# Multipulse Soliton Attractors Facilitated by High-Birefringence Fibers

Xuepeng Wang[1], Haoyu Feng[1], Zhentao Ju[1], Boris A. Malomed[2], and Chaoqing Dai[1*]

[1]*College of Optical, Mechanical and Electrical Engineering, Zhejiang A&F University, Lin'an 311300, China*
[2]*Instituto de Alta Investigación, Universidad de Tarapacá, Casilla 7D, Arica 1000000, Chile*
*Corresponding author email: dcq424@126.com

**The stability of optical solitons is a crucial factor in various applications. This work reveals a novel stable multipulse soliton attractor in fiber lasers. The attractor represents a bound state of multiple solitons, pulling other dynamical states toward itself. By introducing a polarization-maintaining fiber (PMF), the enhanced birefringence in the cavity induces the differential group delay (DGD) between the polarization components of the pulse, leading to soliton splitting and, ultimately, the formation of multi-soliton states. These multi-solitons, affected by intracavity nonlinear and dissipative effects, form attractors consisting of one to seven soliton pulses. Through a combination of experiments and numerical simulations, we systematically investigate the robustness and dynamical behavior of the multipulse soliton attractors. In this context, we analyze disturbances introduced by "rogue" solitons with different frequencies and conduct a comprehensive study of their interaction with the multipulse soliton attractors. The results show that the multipulse soliton attractors are structurally stable and highly resistant to external disturbances, highlighting their potential for high-precision fiber lasers and advanced multipulse systems.**

In recent years, real-time detection techniques[1–3] have advanced the study of soliton[4] dynamics in fiber lasers, extending to multi-wavelength solitons[5], collisions[6–9], and switching behaviors[10,11]. The integration of PMF[12–14], novel saturable absorbers[15], and machine learning methods has significantly enhanced control and complexity in these systems.

Alongside progress in nonstationary regimes[16–20], increasing attention has turned to the formation mechanisms of stable soliton states. Among them, soliton attractors—self-organized, perturbation-resistant states—can return to a fixed pulse shape regardless of initial conditions, making them ideal for high-precision sensing and ultrafast lasers.

Originally proposed four decades ago[21], soliton attractors were mainly explored theoretically via coupled nonlinear Schrödinger equations and related dissipative models[22–24]. More recent experiments have confirmed their existence, including vector[25], polarization[26,27], and multimode attractors[28], laying a foundation for deeper investigation of their stability.

While vector and multimode soliton attractors have been experimentally observed in the context of polarization and modal coupling[25–28], the formation mechanisms and internal dynamics of multipulse soliton attractors in single-mode fiber lasers remain poorly understood. In particular, the role of PMF-induced birefringence in triggering soliton splitting and stabilizing multipulse attractors was not a subject of systematic consideration. While previous studies mostly emphasized formation conditions, our analysis extends to the post-formation stability and interaction dynamics, addressing both temporal and spectral features. We use the phase-space analysis[3,29,30] to characterize the internal dynamics and interaction trajectories of multipulse soliton attractors, which were not addressed in previous experimental works.

As shown in Fig. 1(a), the fiber ring cavity features strong birefringence and is pumped by a 980 nm source via a wavelength-division multiplexer (WDM). The erbium-doped fiber (EDF) absorbs the pump light, excites erbium ions, and generates a 1550 nm output through population inversion. An isolator (ISO) ensures unidirectional transmission, while PC adjusts the polarization state. A panda-type PMF introduces high birefringence, and carbon nanotubes act as the saturable absorber. The cavity totals 14.31 m in length, comprising 12.81 m SMF, 0.45 m EDF, and 1.05 m PMF, with GVD values of -0.0219 $ps^2$/m (SMF, PMF) and 0.0387 $ps^2$/m (EDF). A 7:3 coupler extracts 30% of the output, referred to as the "Total" signal. The cavity operates under pump powers ranging from 150 mW to 160 mW, with multipulse attractors emerging above 150 mW. The CNT-based saturable absorber features the modulation depth of 6.3%, saturation intensity of $\approx$ 0.108 W, and unsaturated loss of $\approx$ 37.4%.

In Fig. 1(a), a PMF-based polarization beam splitter outside the cavity separates the orthogonal polarization components **u** and **v**. By combining oscilloscope traces and spectral data, we identify these as polarization-rotating vector solitons, characterized by amplitude and frequency coupling. The components remain phase-locked via cross-phase modulation (XPM), and their polarization states undergo periodic rotation due to the combined effects of birefringence and nonlinearity, conserving energy. The apparent period doubling between components $u$ and $v$ arises from the polarization-rotation dynamics of the vector solitons, where two orthogonal components $u$ and $v$ exhibit the complementary evolution, resulting in spectral modulation patterns with effectively doubled periods. The Stokes evolution confirms continuous polarization rotation, with Fig. 1(b) showing initial (blue) and final (green) positions along a closed trajectory. The green point's central location suggests repeated oscillations in a cyclic manner.

The complementary nature of the two components ensures that a single polarization channel reflects the soliton's essential features—such as pulse width, spectral profile, and dynamics. Thus, analyzing one component suffices to capture the behavior of the polarization-rotating vector soliton.

By tuning intracavity power, we observe the emergence of multipulse soliton attractors. Regardless of initial conditions, the system evolves toward a stable attractor. Notably, soliton amplitudes converge to a fixed value per round trip, independent of pulse number[22]. As shown in Fig. 1(c), these attractors exhibit uniform amplitude, pulse width, and velocity, forming a robust localized state.

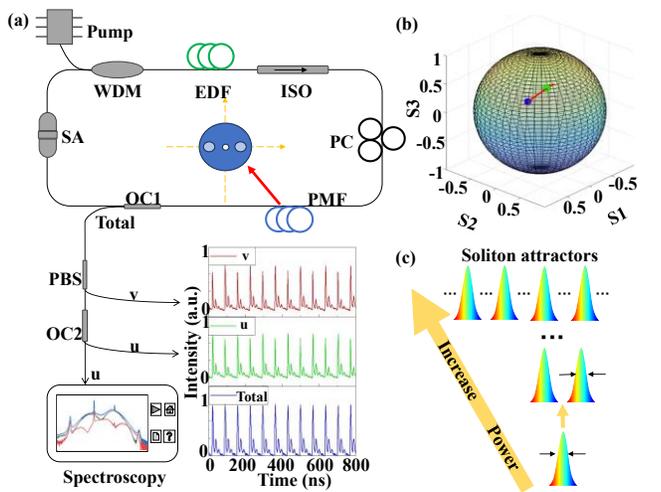

Fig. 1. The experiment setup and principles of its operation. (a) The device setup for the simulations and experiment. (b) The Stokes distribution of the polarization-rotating vector solitons. (c) The schematical diagram of the evolution of the soliton attractors.

In the experiment, a 1.05-meter high-birefringence PMF induces DGD, splitting the initial pulse into orthogonal polarization components. The phase difference is tuned via the PC, introducing the initial symmetry breaking for multi-soliton formation. Dissipative effects—such as net positive dispersion and cavity loss—drive phase-space contraction, selecting stable amplitude–width pairs. Meanwhile, nonlinear interactions, particularly XPM, enable energy exchange between components, establishing soliton coupling. The interplay between birefringence-induced asymmetry and dissipative regulation ensures convergence to a stable multipulse soliton attractor, regardless of initial conditions, after several cavity round-trips.

The system is modeled by coupled generalized nonlinear Schrödinger equations (GNLSE) incorporating DGD effects.

Fig. 2(a) presents the distribution of soliton attractors with varying pulse numbers as a function of small-signal gain $g_0$ and gain saturation energy $E_{sat}$. Figs. S1(a–g) illustrate the simulated evolution of polarization component u, showing stable multipulse states from 1 to 7 pulses. The introduction of polarization group delay via the PMF facilitates soliton formation. Under the interplay of dissipation and nonlinearity, energy exchange between solitons suppresses nonlinear phase accumulation. As the system evolves, the relative velocities between pulses diminish and eventually vanish, indicating the formation of a stable soliton attractor.

Fig. 2(b) shows the evolution of vector solitons obtained from simulations of Eqs. (1), (2), where the "vector" nature refers to the two-component polarization structure of the multipulse state. Nine solitons with varying initial velocities and amplitudes converge to a common fixed point, forming a stable soliton attractor characterized by equalized amplitudes (~0.75) and vanishing relative velocities.

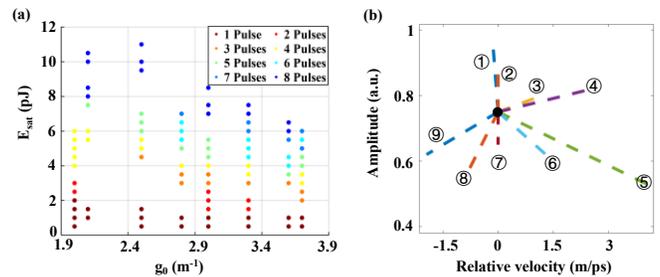

Fig. 2. Simulation results of the solitons evolution. (a) The plot showing the number of pulses in the states that may serve as the soliton attractors, for different values of $E_{sat}$ and $g_0$. (b) Evolution of solitons ①–⑨ toward the formation of a stable multipulse soliton attractors for initial relative velocities -0.14, 0, 1.14, 2.43, 3.86, 1.43, 0, -0.86, -1.86(m/ps) and amplitudes 0.92, 0.87, 0.8, 0.82, 0.53, 0.6, 0.65, 0.55, 0.62(a.u.), respectively.

Further verification of the soliton attractor is provided in Fig. 3, which compares the pulse widths and amplitudes from both simulations and experiments. The identification of attractor characteristics—such as uniform amplitude and pulse width—under varying pulse numbers is presented in Supplement 1.

Parallel to the simulation results in Fig. S1(d), the experimentally measured spectrum of the four-pulse soliton attractor is shown in Fig. 3(a), with a spectral width of 2.8 nm. Using dispersion-stretching techniques, the soliton dynamics is captured in Fig. 3(b), with individual Pulses (I-IV) identified. Their periodic structure is shown in Fig. 3(d), where Pulses I-III are relatively separated, while Pulses III and IV are closely spaced.

Fig. 3(c) presents the relative phases $\theta_{(3,1)}$, $\theta_{(3,2)}$, and $\theta_{(3,4)}$, which fluctuate within 25 rad, with respective time intervals of 36 ns, 13 ns, and 6 ns. The interaction plane between Pulses III and II in Fig. 3(d) shows a semicircular arc with a radius equal to their time delay. Combined with the 3D trajectory in Fig. 3(e), the data demonstrate periodic oscillations in the relative phase. The fixed time intervals lead to synchronized spectral fringes, while the interaction plane reveals circular oscillations. These periodic phase evolutions indicate that the weak interactions between pulses remain in a balanced state, allowing the relative phases to oscillate back and forth stably over time.

To further study properties of the soliton attractor, the five-pulse form was experimentally captured, corresponding to the simulations presented in Fig. S1(e). The spectral data of the soliton attractor is shown in Fig. 4(b), with the spectral width of 1.1 nm.

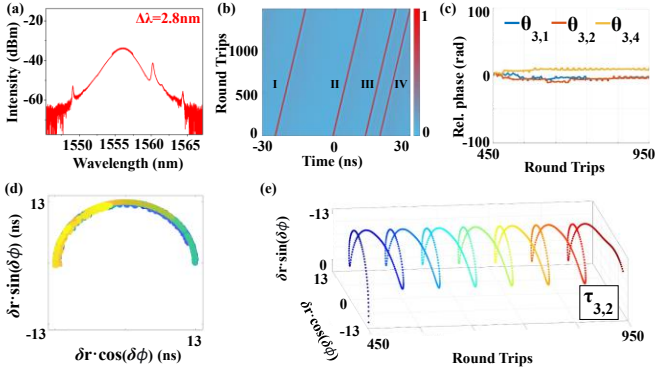

Fig. 3. Experiment results for the soliton attractor with the four-pulse form. (a) The spectrum and (b) real-time dynamical evolution. Panel (c) shows the relative phases, (d) shows the 2D picture of the interaction between Pulse III and the other ones, and (e) shows the 3D picture of the interaction between Pulses III and II.

To fully investigate the stability and internal features of the attractor state, we analyzed the relative phases and trajectories pf the interaction between individual pulses. To clearly visualize these dynamics, Fig. 4(c) presents representative 3D trajectories between Pulse V and Pulses II and IV, projected onto a semi-cylindrical surface where the radius corresponds to the temporal interval in each pulse pair. The interaction between Pulses V and II exhibits periodic oscillations along the circular arc, indicating dynamical binding. However, the trajectory between Pulses V and IV evolves smoothly along the arc without noticeable perturbations, suggesting the strong phase synchronization. The interactions between Pulse V and Pulses I and III are similar to the oscillatory trajectories of the V–II pair, further confirming the strong binding among these pulses.

For the comprehensive study of the multipulse soliton attractor, the relative phase and interaction trajectories for Pulses II and III were further analyzed. As shown in the inset to Figs. 4(c) and S3(d), their relative phase exhibits oscillations with a time interval of 18 ns. The 3D trajectory analysis reveals that their evolution features small-amplitude periodic oscillations along a circle with the time interval as the radius, indicating a state with weak coupling between them.

The stability of the soliton attractor against external interference is experimentally demonstrated in Fig. 5(a) corresponding to the simulations displayed above in Fig. S1(g). Unlike the two single-wavelength cases shown in Figs. 3 and 4, by adjusting PC, the polarization direction of the pulses inside the cavity is set at a specific angle relative to the fast/slow axes of the PMF, effectively separating the two orthogonal polarization modes and accumulating the phase difference. The high birefringence of the PMF, along with its length, determines the free spectral range, $FSR = \lambda^2/(\Delta n \cdot L)$. The periodic transmission peaks of the PMF, when superimposed with the spectrum of the erbium-doped fiber amplifier, interact with the polarization-dependent loss introduced by the PC, suppressing the mode competition and producing the filtering effect[31–33]. When the PC is adjusted to a specific state, the loss difference between the two wavelengths is balanced, allowing the resonance condition to be satisfied simultaneously, enabling the dual-wavelength co-propagation. In the cavity with the seven-pulse soliton attractor state, this filtering effect generates new frequency components inside the cavity. As shown in Fig. 5(a), stable Pulses I-VII form a multipulse soliton attractor, while Pulse VIII is a free "rogue" pulse with different frequency components. Due to the differences in group velocity corresponding to the various frequency components, pulses with different group velocities collide with each other.

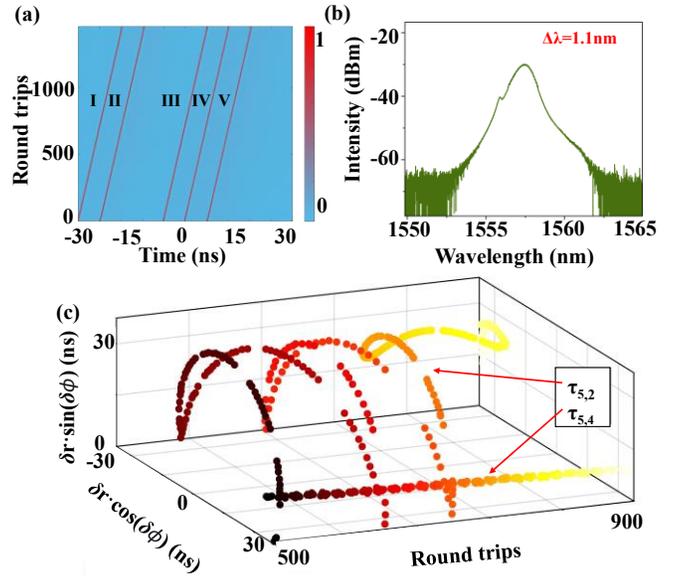

Fig. 4. Experimental results for the five-pulse soliton attractor. (a) The real-time evolution diagram. (b) The spectrum and (c) 3D trajectories of the interaction between Pulse VI and Pulses II and IV.

The respective spectral data are shown in Fig. 5(b), where the new pulse with the frequency around 1560 nm appears, in addition to the original 1555 nm component, resulting in a dual-wavelength spectrum. Under the action of the extracavity attenuator, the pulse corresponding to the low-frequency component disappears, and only the high-frequency component of the stabilized soliton attractor persists, which matches the spectral analysis shown in Fig. 5(b).

As shown in Fig. 5(c), the energy of Pulse V temporarily fluctuates during the collision, but subsequently returns to its original level, indicating that the attractor structure remains stable and unaffected in the long term.

Finally, the interaction between the "rogue" Pulse VIII and Pulse V was analyzed. Unlike the four-pulse and five-pulse soliton attractors in Figs. 3(e), 4(c) and S3(d), Pulse VIII has a different group velocity. This causes the time interval between the solitons to gradually decrease, and after the collision, the interval slowly increases. The projection in Fig. 5(c) shows the trajectory as it spirals around the origin. The two solitons collide at around the 3500th RT, where the time interval is zero, and after that, the radius gradually increases, moving outward in a sliding motion. The trajectory plot and interval graph exhibit a sliding phase form, with no irregular trajectories observed around it. This suggests that the collision does not significantly affect the stability of the solitons, and the continuity of their internal phase information is not disrupted. This observation further confirms that the soliton attractors not only resist random disturbances, but also withstand collisions with "rogue solitons." These findings provide strong evidence for the inherent robustness of the multipulse soliton attractors against external perturbations and internal collisions alike.

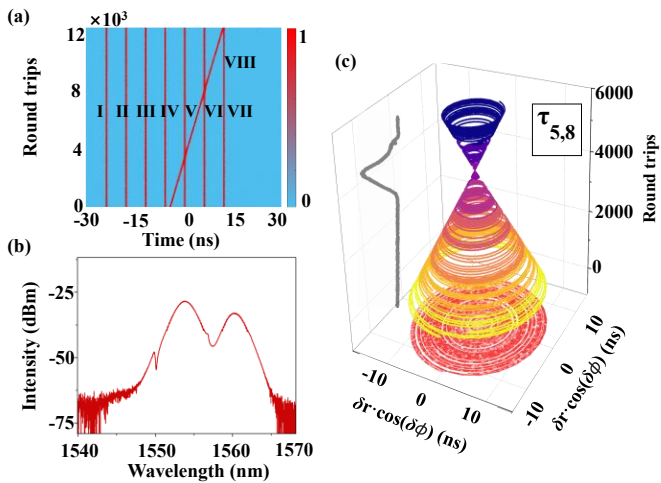

Fig. 5. Experiment results for the collisions of freely moving "rogue" solitons with the attractor in the form of the seven-pulse complex. (a) The collision scenario; (b) the spectrum and (c) the 3D interaction diagram between Pulse VIII and Pulse V, and the energy evolution of Pulse V (the gray curve).

By incorporating highly birefringent fibers to construct coupled dissipative systems, this study reveals the formation mechanism and stability characteristics of multipulse soliton attractors. The PMF-induced DGD initiates pulse splitting, while nonlinear and dissipative effects guide sub-pulses to converge to uniform amplitude, width, and velocity, forming a stable multipulse attractor. Beyond time-domain and energy-distribution analyses, phase-space methods uncover sliding and oscillatory phase trajectories, offering an alternative verification of stability and complementing spectral and theoretical approaches. The use of PMF enables robust manipulation of attractor states and facilitates controlled collisions via group velocity differences, thereby evaluating anti-interference performance.

These results confirm the dual robustness of multipulse soliton attractors: internal structural stability and resilience to external perturbations. The findings provide new insights into soliton interaction dynamics and establish a foundation for optimizing fiber laser systems, with potential applications in ultrafast lasers and high-precision sensing.

**Funding.** National Natural Science Foundation of China (Grant Nos. 12261131495 and 12475008); Scientific Research and Developed Fund of Zhejiang A&F University (Grant No. 2021FR0009); Israel Science Foundation (Grant No. 1695/22).

**Disclosures.** The authors declare no conflicts of interest.

**Data availability.** Data underlying the results presented in this paper are not publicly available at this time but may be obtained from the authors upon reasonable request.

**Supplemental document.** See Supplement 1 for supporting content.